\documentclass[prb,aps,twocolumn]{revtex4}
\usepackage{amsmath}
\usepackage[english]{babel}
\usepackage{graphicx}

\begin{document}

\title{Tunneling conductance due to discrete spectrum of Andreev states}

\author{P.\ A.\ Ioselevich}
\affiliation{L.\ D.\ Landau Institute for Theoretical Physics, Kosygin str.\ 2,
Moscow, 119334 Russia}
\affiliation{Moscow Institute of Physics and Technology, Institutsky per.\ 9,
Dolgoprudny, 141700 Russia}

\author{M.\ V.\ Feigel'man} 
\affiliation{L.\ D.\ Landau Institute for Theoretical Physics, Kosygin str.\ 2,
Moscow, 119334 Russia}
\affiliation{Moscow Institute of Physics and Technology, Institutsky per.\ 9,
Dolgoprudny, 141700 Russia}

\begin{abstract}
We study tunneling spectroscopy of discrete subgap Andreev states in a superconducting system. If the tunneling coupling is weak, individual levels are resolved and the conductance $G(V)$ at small temperatures is  composed of a set of resonant Lorentz peaks which cannot be described within perturbation theory over tunnelling strength. 
We establish a general formula for their widths and heights and show that the width of any peak scales as normal-state tunnel conductance, while its height is $\lesssim 2e^2/h$ and depends only on contact geometry and the spatial profile of the resonant Andreev level. We also establish an exact formula for the single-channel conductance that takes the whole Andreev spectrum into account. We use it to study the interference of Andreev reflection processes through different levels. 
The effect is most pronounced at low voltages, where an Andreev level at $E_j$ and its conjugate at $-E_j$ interfere destructively. 
This interference leads to the quantization of the zero-bias conductance: $G(0)$ equals $2e^2/h$ (or $0$) if there is (there is not) a Majorana fermion in the spectrum, in agreement with previous results from $S$-matrix theory. 
We also study $G(eV>0)$ for a system with a pair of almost decoupled Majorana fermions with splitting $E_0$ and show that at lowest $E_0$ there is a zero-bias Lorentz peak of width $W$ as expected for a single Majorana fermion (a topological NS-junction) with a narrow dip of width $E_0^2/W$ at zero bias, which ensures $G(0)=0$ (the NS-junction remains trivial on a very small energy scale). As the coupling $W$ gets stronger, the dip becomes narrower, which can be understood as enhanced decoupling of the remote Majorana fermion. Then the zero-bias dip requires  extremely low temperatures $T\lesssim E_0^2/W$ to be observed.
\end{abstract}

\maketitle

\section{Introduction}
Scanning Tunneling Microscopy (STM) is a powerful method of studying electronic surface properties of condensed matter systems. Electrons from the metallic tip of the microscope tunnel into the system, creating a stationary current at fixed voltage. At low temperatures the conductance $G$ of the tunnelling contact is proportional to the surface density of states $\rho$  beneath the tip, and the coupling strength between the tip and the system 
\begin{equation}
G(V)\propto |t^2|\rho(r,eV).\label{didvclassic}
\end{equation}
This relation allows to probe local density of states with energy and spatial resolution. In recent works \cite{FuKane4pi,LeeNg,Sarma,OregOppen,Flensberg,IoselFeig2011,OppenShapiro,IOF,Sarma2012,Magneto4pi,SarmaNanowires,PikulinBeenakker,Aguado,Aguado2} transport measurements have been suggested to probe the Majorana fermion in superconducting structures with non-trivial topological properties, such as nanowires combining strong spin-orbit, induced superconductivity and magnetic field \cite{SarmaNanowires,OregOppen,Sarma}, and vortex cores on topological insulator surfaces with induced s-wave superconductivity \cite{FuKane2008,Nori,IoselFeig2011,IOF}. There are two major transport phenomena suggested, the first one is the fractional Josephson effect \cite{FuKane4pi,IoselFeig2011,Magneto4pi,Beenakker4pi} : the current-phase relation in an SNS-junction hosting a pair of Majorana fermions is $4\pi$-periodic with phase if the fermionic parity in the junction is conserved \cite{Kitaev}. This effect also implies the doubling of Shapiro steps in an \textit{ac} measurement \cite{OppenShapiro}, which has recently been reported \cite{Rokhinson}. A second, and more direct effect of the Majorana state on transport is the peak in zero-bias conductance. Transport experiments have been done on nanowire-based setups \cite{Mourik,Heiblum,LundGroup}, showing a zero-bias peak, that is robust to the change of magnetic field (in both magnitude and direction) and gate voltages. 
Under the conditions of these experiments, the Majorana state is localized and thus is expected to produce a conductance peak proportional to the peak in the density of states,
in agreement with Eq.\eqref{didvclassic}.
For the vortex core, direct STM measurements have been proposed \cite{Nori,IOF}  with a similar prediction of a zero-bias conductance peak.

The idea of probing the Majorana bound state with tunneling transport measurements motivated us to revisit the conventional relation \eqref{didvclassic} for the case of a Majorana level and, more generally, for an arbitrary discrete spectrum of Andreev bound states. In this work we study the conductance through a tunneling contact between a normal metal and a superconductor hosting a subgap spectrum of Andreev bound states $E_j$. While the $G(V)$ characteristic is expected to reflect the structure of the spectrum and exhibit peaks at voltages close to $E_j/e$, it is clear, that the classical expression \eqref{didvclassic} cannot accurately describe the current in our system of interest. Since we consider a discrete spectrum, the density of states $\rho(E)$ is a set of $\delta$-functions. The conductance however, cannot be a set of $\delta$-functions as it is limited by the number of channels in the contact. Moreover, expression \eqref{didvclassic} is a Fermi Golden rule result, that describes electrons slowly leaking the probed system. In our superconducting system there are no extended states at subgap energies and single electrons cannot leak into it. The only possibility for a stationary current is due to Andreev reflection: the process of electrons being reflected from the NS-junction as holes. The discrete Andreev levels in the superconducting system act as quasi-stationary levels for this process: an incident electron may occupy an energetically favorable level and is then emitted back into the metal tip either as an electron or as a hole. At voltages close to one of the levels $E_{j_0}$ this process should exhibit resonant behavior, and cannot be described perturbatively in terms of the coupling strength. Our aim is to describe these resonances for a generic tunneling contact.

Another interesting problem of the discrete spectrum is interference of Andreev reflection processes involving different levels. There is no interference involved in the simple formula \eqref{didvclassic}, where the conductance is directly proportional to the number of available energy levels on the probed surface. The process of Andreev reflection is different from conventional tunneling as it involves tunneling into the probed system and then back into the metal. Reflection of the electron into a certain hole channel can happen in multiple ways, that differ in the Andreev level $E_j$ they tunnel into. The amplitudes of these different ways have to be summed together, which allows for interference between different levels. The interference effects should be important for a few-channel contact, especially a single-channel one, where all amplitudes are added together. Most interesting is the case of low voltages, where the BdG-symmetry of the spectrum becomes important and defines the interference between conjugate levels $\pm E_j$. 

In section \ref{sec:setup} we formally introduce the system we study and the formalism we will use. This includes the Landauer formula for superconductors, a Kubo-type formula for Andreev reflection, and the $\check{T}$-matrix expression of exact Green's functions in terms of the tunneling Hamiltonian and unperturbed Green's functions of the connected systems. In section \ref{sec:singlelevel} we study the resonance peaks that occur at voltages close to a single Andreev level $E_{j_0}$. We derive general formulae describing the peaks and study the particular cases of a peak on a Majorana level, formulae for a point-contact and the conductance at finite temperatures. In section \ref{sec:singlechannel} we establish an exact formula for conductance in the single-channel contact case that takes the whole spectrum $E_j$ into account. We then study interference of different levels and observe quantization of zero-bias conductance. Next, we  study conductance at finite voltages. In particular, we consider a system with a pair of weakly coupled Majorana levels. We find that a zero-bias peak in this system coexists with an extremely narrow dip at $E=0$ that shows that the system is still topologically trivial. We discuss the experimental requirements to observe the dip structure of this peak and show that, surprisingly, they become more severe when the tunnelling coupling to the probe increases. 
In section \ref{sec:discussion} we discuss some experimentally realized systems where level spacing of Andreev states is large enough for the individual levels to be observed.

\section{Setup}\label{sec:setup}
The system we are studying consists of a metallic tip described by the Hamiltonian $H_M$, a superconducting system $H_S$ and a tunneling Hamiltonian $H_T$ that allows electrons to tunnel from the tip onto the system.
\begin{equation}
H=H_M+H_T+H_S
\end{equation}
We are interested in voltages and temperatures much lower than the superconducting gap and thus only take the Andreev levels into account:
\begin{equation}
H_S=\sum_jE_j|j\rangle\langle j|\label{HS}
\end{equation} 
The contact is described by the tunnelling Hamiltonian
\begin{equation}
H_T=\sum\limits_{m,s}|\theta_m\rangle t_{ms}\langle\sigma_s| + h.c.
\end{equation}
where $|\theta_m\rangle,|\sigma_s\rangle$ describe some electronic wave functions in the metallic tip and superconductor, correspondingly. In the Bogolyubov-de Gennes formalism we will use, $H_T$ has to be extended into Nambu space
\begin{multline}
\check{H}_T=\begin{pmatrix}H_T&0\\0&-\hat{\Theta}H_T\hat{\Theta}^{-1}\end{pmatrix}=\\
\sum\limits_{m,s}|\theta_m\rangle t_{ms}\langle\sigma_s|-|\theta^*_m\rangle t^*_{ms}\langle\sigma^*_s|+h.c.
\end{multline}
where $\hat{\Theta}$ is the time-reversal operator, and $|\theta^*_m\rangle$ is the particle-hole--conjugated partner of $|\theta_m\rangle$, i.e. $|\theta^*_m\rangle=\mathcal{C}|\theta_m\rangle$ with $\mathcal{C}$ denoting particle-hole conjugation. Throughout the paper the check sign indicates a structure in Nambu space. 

At subgap energies $E<\Delta$ electrons incident from the metallic tip are reflected back either normally or as holes. The latter process is Andreev reflection and produces a current through the NS-junction, which can be described in the Landauer formalism. The formula describing the stationary current at some voltage bias reads (we derive it from the general Lesovik-Levitov formula \cite{LevitovLesovik} for the BdG-Hamiltonian)  
\begin{equation}
I=\frac{2e}{h}\int\limits_0^\infty T_{A}(E){\big[} f_0(E-eV)-f_0(E+eV){\big]}dE\label{levitov}
\end{equation}
Note that the factor of $2e$ is due to the quantum of transferred charge in the Andreev process.
 $f_0(E)$ denotes the equilibrium Fermi distribution function, and $T_{A}(E)$ is the sum of Andreev reflection eigenvalues:
\begin{align}
T_A(E)=\mathrm{tr\phantom{,}} \left(S_{eh}(E)^\dagger S_{eh}(E)\right)\label{TAdefinition}
\end{align}
where $S_{eh}(E)$ is the electron-hole block of the $S$-matrix of the N-S junction. With the help of the relations $T_A(E)=T_A(-E)$ and $f_0(E)+f_0(-E)=1$ formula \eqref{levitov} can be rewritten in a form similar to the one for the standard N-N contact current:
\begin{align}
I=\frac{2e}{h}\int\limits_{-\infty}^\infty T_{A}(E){\big[} f_0(E-eV)-f_0(E){\big]}dE\label{levitov2}\\
G=\frac{dI}{dV}=\frac{2e^2}{h}\int\limits_{-\infty}^\infty  \frac{T_{A}(E)}{4T\cosh^2\left(\frac{E-eV}{2T}\right)}dE\label{levitovG}
\end{align}
Below we mainly study the zero-temperature conductance $G(eV)=T_A(E=eV)\cdot2e^2/h$ which is just the sum of Andreev transmission eigenvalues times the relevant conductance quantum $2e^2/h$. Conductance at $T>0$ can then be easily obtained from the zero-temperature result via the convolution \eqref{levitovG}.
The trace \eqref{TAdefinition} can be written as
\begin{align}
T_A=\mathrm{tr\phantom{,}}\left(\begin{pmatrix}\hat{v}&0\\0&0\end{pmatrix}\check{G}_E^A\begin{pmatrix}0&0\\0&\hat{v}^*\end{pmatrix}\check{G}_E^R\right)\label{Kubo}
\end{align}
which can be understood as an Andreev-reflection analog of the Kubo formula for the linear response.
 $\hat{v}$ is the velocity operator for electrons and $\hat{v}^*=\hat{\Theta}\hat{v}\hat{\Theta}^{-1}$ is its time-reversal. The explicit matrices inside the trace \eqref{Kubo} act in Nambu space, and $\check{G}^{A,R}_E$ are exact advanced and retarded Green's functions at energy $E$. The exact Green's functions in the metal are exressed through unperturbed Green's functions $\check{G}_0=[E+i0-\check{H}_M]^{-1}$ and $\check{G}_S=[E+i0-\check{H}_S]^{-1}$ (here and in what follows Green's functions without index are assumed retarded):
\begin{align}
\check{G}=\check{G}_0+\check{T}\check{G}_0\label{G}\\
\check{T}=\frac{\check{G}_0\check{H}_T\check{G}_S\check{H}_T}{1-\check{G}_0\check{H}_T\check{G}_S\check{H}_T}\label{Tmatrix}
\end{align}
The Green's functions here act in the Nambu space, so 
\begin{align}
\check{G}_0=\begin{pmatrix}E+i0-H_M & 0\\ 0 & E+i0+H_M^* \end{pmatrix}^{-1}=\begin{pmatrix}g_E & 0\\ 0 & -g^*_{-E}\end{pmatrix}
\end{align}
We use $g$ to denote normal metal Green's functions without Nambu structure and $g^*=\hat{\Theta}g\hat{\Theta}^{-1}$ for its time-reversal image. The superconducting Green's function is
\begin{align}
\check{G}_S=\sum\limits_{j>0}\frac{|j\rangle \langle j|}{E+i0-E_j}+\frac{|j^*\rangle \langle j^*|}{E+i0+E_j}\label{GS}
\end{align} 
The sum over $j>0$ reflects the symmetry of the levels -- each positive level $|j\rangle $ has a particle-hole conjugated partner $|j^*\rangle$ with opposite energy. $G_S$ has poles at $E=\pm E_j$ and the $\check{T}$-matrix cannot be treated perturbatively in their vicinity. All terms of the geometrical sequence have to be included in the expression \eqref{Tmatrix}.  This can be done in a compact way in two cases we consider below.

\section{Single-level resonance for arbitrary $H_T$}\label{sec:singlelevel}
The $\check{T}$-matrix and consequently $T_A$ can be found explicitly in a closed form in the resonance situation $|E-E_{j_0}|\ll |E-E_{j\neq j_0}|$ when the energy of the incident electron is closer to the Andreev level $E_{j_0}$ than to any other level $E_j$. We only keep the $(E-E_{j_0})^{-1}$ term in $\check{G}_S$ in this case, and write
\begin{align}
\check{H}_T\check{G}_S\check{H}_T=\frac{\check{G}_0|\tau\rangle\langle\tau|}{E-E_{j_0}+i0}\qquad\mbox{with}\quad |\tau\rangle=\check{H}_T|j_0\rangle\label{tau}
\end{align}
Here $|\tau\rangle$ is a function of the electron coordinates in the metallic tip, which is determined by the contact Hamiltonian $H_T$ and the resonant Andreev state $|j_0\rangle$. It has both electron and hole components which we denote as $|\tau_e\rangle$ and $|\tau_h\rangle$ for later use. The simple projector form of the operator \eqref{tau} leads to
\begin{align}
\check{T}=\frac{\check{G}_0|\tau\rangle\langle\tau|}{E-E_{j_0}+i0-\langle\tau|\check{G}_0|\tau\rangle}\label{Tmatrixres}
\end{align}
The real part of $\langle\tau|\check{G}_0|\tau\rangle$ is zero, and we denote its imaginary part as $W$.
Substituting the $\check{T}$-matrix \eqref{Tmatrixres} into formula \eqref{G} and the result into the Kubo formula \eqref{Kubo}, we find 
\begin{align}
G=\frac{2e^2}{h}T_A=\frac{\langle\tau_e|g^A_E\hat{v}g^R_E|\tau_e\rangle\langle\tau_h|g^{R*}_{-E}\hat{v}^*g^{A*}_{-E}|\tau_h\rangle}{(E-E_{j_0})^2+W^2}\\
W=\pi(n_e+n_h)\quad\mbox{with}\quad
\begin{cases}
n_e=\langle\tau_e|\frac{g^R_E-g^A_E}{2\pi i}|\tau_e\rangle\\
n_h=\langle\tau^*_h|\frac{g^R_{-E}-g^A_{-E}}{2\pi i}|\tau^*_h\rangle
\end{cases}\label{W} 
\end{align}
(here $|\tau_h^*\rangle=\hat{\Theta}|\tau_h\rangle$ is just the time-reversed wave function)
In the case of a clean metallic tip (and assuming that Fermi energy $E_F$ is the largest energy scale), the $gvg$-terms can be simplified in terms of  the  $n_{e,h}$  densities: 
\begin{align}
G(eV)=\frac{2e^2}{h}\frac{1}{1+\frac{(eV-E_{j_0})^2}{W^2}}T_A^*\label{Gres}\\
T_A^*=\frac{4n_en_h}{(n_e+n_h)^2}
\end{align}
Formula \eqref{Gres} constitutes our main result for the single-level resonance, and we will now discuss it in more detail. The $V$-dependence of \eqref{Gres} has a Lorentz shape, which is generic: at resonance, the $S$-matrix is governed by the pole at $E=E_{j_0}+iW$. The width $W$ describes the broadening of the Andreev level at the energy $E_{j_0}$ due to coupling to the tunneling probe. $T_A^*\leq1$ is the probability of Andreev reflection for an incident electron with $E=E_{j_0}$. It is defined by the relation between the electron and hole components of $|j_0\rangle$ and the geometry of the contact. If $|j_0\rangle$ was a purely electronic state $T_A^*\sim n_h=0$ and there would be no Andreev reflection from $|j_0\rangle$ and hence no conductance. For deep subgap states ($E_0 \ll \Delta$)  electron and hole components are very similar (e.g. connected by phase rotation for a uniform $\Delta(r)$) so we expect $1-T_A^*\ll 1$. However, such a feature  may be absent when measuring tunneling conductance at points  where the gap $\Delta$ is locally suppressed, $\Delta(r) \leq E_0$, e.g. near the origin of a vortex core, or in the middle of an SNS-junction. In such regions electron and hole components may differ considerably, leading to a small $T_A^*$. The presence of the $T_A^*$ factor in \eqref{Gres} is an important difference from the classical relation \eqref{didvclassic}: Andreev reflection amplitude depends on the $e-h$ structure of the wave function, and not just on the overall probability density.

\subsection{Majorana peak}
An important special case of the single-level resonance described by \eqref{Gres} is the resonance on a Majorana bound state. In this case $|j_0\rangle$ is self-conjugate, so that $|\tau_e\rangle=|\tau^*_h\rangle$. Hence $n_e=n_h$ and $T_A^*=1$,
\begin{align}
G=\frac{2e^2}{h}\frac{1}{1+\frac{(eV)^2}{W^2}}\label{GresMajo}
\end{align}
This agrees with earlier derivations of the Majorana-induced zero-bias tunneling conductance peak \cite{LeeNg,Flensberg}. The width $W$ of the peak still depends on contact and wave function geometry, while the height is fixed to $2e^2/h$ by the symmetry of the Majorana state. Other energy levels $|j\neq j_0\rangle$ generally lead to a negative correction to the conductance of the order of $O(W^2/\omega_0^2)$ where $\omega_0$ is the typical level spacing of the Andreev spectrum in Eq.\eqref{HS}.

\subsection{Point-contact}
The case most relevant to a scanning tunneling microscope setup is the point contact $H_T=|\theta\rangle t_0\langle\sigma|+h.c.$ where wave functions $|\theta\rangle $ and $|\sigma\rangle $ are localized at some points $r_M$ and $r_S$ in the metallic tip and superconductor correspondingly. For simplicity we assume here that $H_T$ commutes with spin, but our results are easily generalized to a spin-sensitive setup.
\begin{align}
T_A^*=4\frac{|\psi_e^2||\psi_h^2|}{\left(|\psi_e^2|+|\psi_h^2|\right)^2}\label{Wpointlike1}\\
W=\pi|t_0|^2\nu_M|\psi|^2 \label{Wpointlike2}
\end{align}
Here $\nu_M$ is the density of states in the metallic tip per spin projection and $|\psi_e(h)^2|=|\psi_{e(h)\uparrow}^2|+|\psi_{e(h)\downarrow}^2|$ is the probability density of the electron (hole) component of the resonant Andreev level. We can rewrite \eqref{GresMajo} for the point-contact as
\begin{align}
G(eV)=4\pi^2|t_0^2|\frac{e^2}{h}\nu_M\tilde{\nu}T_A^*,\label{Gquasinormal}
\end{align}
which has the form of a standard expression for normal conductance times the resonant reflection probability $T_A^*$. $\tilde{\nu}$ in Eq.\ \eqref{Gquasinormal} is a broadened density of states of the Andreev level $|j_0\rangle$:
\begin{align}
\tilde{\nu}(eV)=\frac{|\psi^2|}{2}\cdot\frac{1/(\pi W)}{1+(eV-E_{j_0})^2/W^2}\label{nubroadened}
\end{align}
The point-contact expression \eqref{Gquasinormal} can be generalized to an arbitrary contact:
\begin{align}
G=-\frac{e^2}{2h}\mathrm{tr\phantom{,}}\left(\check{H}_T\left[\check{G}^R_0-\check{G}^A_0\right]\check{H}_T\left[\check{\tilde{G}}^R_S-\check{\tilde{G}}^A_S\right]\right)T_A^*\label{Gquasinormal2}
\end{align}
where $\check{\tilde{G}}_{S}$ is the broadened Green's function of the probed system:
\begin{align}
\check{\tilde{G}}(eV)=\int \frac{1/(\pi W)}{1+(\epsilon-eV)^2/W^2}\check{G}(\epsilon)d\epsilon\label{Gbroadened}
\end{align}
If instead of $\check{\tilde{G}}_S$  in \eqref{Gquasinormal2} we substitute the Green's function corresponding to the normal state of our probed system, and put $T_A^*=1$, the expression \eqref{Gquasinormal} will yield the normal-state conductance. In the superconducting regime in presence  of the Andreev spectrum, the Green's function pole has to be broadened according to Eq.\ \eqref{Gbroadened}, so that $\delta$-peaks in $G^R-G^A$ are smeared into Lorentz peaks of the width $W$. We remind that  the parameter $W$ is different for different peaks according to Eq.\ \eqref{W}. Thus, if $W$ is calculated for $|j_0\rangle$, then expression \eqref{Gbroadened} with that $W$ only works for the peak at $eV=E_{j_0}$.

\subsection{Temperature dependence}
Given the zero-temperature $G(T=0,eV)$ from \eqref{Gres}, the finite-temperature conductance is readily obtained via formula \eqref{levitovG} by convoluting it with the thermal distribution  function. The shape of $G(T)$ starts to differ from the $G(T=0)$ case as soon as $T$ becomes comparable to $W$. At this point the peaks become broader, while retaining spectral weight. At $\omega_0\gg T\gg W$ (we denote the level spacing by $\omega_0$) the temperature-broadened resonant peak looks like
\begin{align}
G(eV)=\frac{2e^2}{h}T_A^*\frac{\pi W}{T\cosh^2\left(\frac{E_{j_0}-eV}{2T}\right)},\qquad W\ll T\ll \omega_0 
\end{align}
Here $W$ enters  as the spectral weight of the original Lorentz peak \eqref{Gres}, while the peak shape is defined by the temperature $T$ only. If instead of expression \eqref{Gres} we use expression \eqref{Gquasinormal} or \eqref{Gquasinormal2} in the convolution \eqref{levitovG},  we find that the broadening of the $\delta$-peaks by $W$ is much less than thermal broadening and thus drops out. For the point-contact 
\begin{align}
G(eV)= T_A^*\pi^2\frac{e^2}{h}\frac{|t_0^2|\nu|\psi^2|/2}{T\cosh^2\left(\frac{E_0-eV}{2T}\right)},\qquad W\ll T\ll \omega_0. \label{GT}
\end{align}

\section{Single-channel contact, multi-level system}
\label{sec:singlechannel}
 Another case which can be solved exactly is presented by  a single-channel contact $H_T$ coupled to  an arbitrary system of Andreev levels $H_S$. In this section we account for all Andreev levels simultaneously in an exact way, which allows us to study important interference effects due to Andreev reflection via different levels of the discrete spectrum. 

An arbitrary single-channel tunneling Hamiltonian can be written as
$h_T=|\theta\rangle t\langle\sigma|+h.c.$ with a real $t$ and normalized electronic wave functions $|\theta\rangle $ and $|\sigma\rangle $ that are defined in the metal and superconductor respectively. To include $h_T$ into formula \eqref{Tmatrix}, we extend it into Nambu space:
\begin{align}
\check{H}_T=|\theta\rangle t\langle\sigma|-|\theta^*\rangle t^*\langle\sigma^*|+h.c.\label{htscs}
\end{align}
Thus, in the BdG formalism $\check{H}_T$ appears as a two-channel contact. The calculation of the $\check{T}$-matrix is similar to that of the expression \eqref{Tmatrixres}. Using \eqref{htscs} and \eqref{GS} we write
\begin{align}
\check{G}_0\check{H}_T\check{G}_S\check{H}_T=\check{G}_0\begin{pmatrix}|\theta\rangle & |\theta^*\rangle\end{pmatrix}
N_0^{-1}\check{\Sigma}
\begin{pmatrix}\langle\theta|\\ \langle\theta^*|\end{pmatrix}\label{T1}
\end{align}
with a $2\times2$ matrix $\check\Sigma$
\begin{align}
\check{\Sigma}=\begin{pmatrix}\Sigma_e & \Sigma_A \\ \Sigma_A^* & \Sigma_h
\end{pmatrix}\\\label{Sigmacheck}
\Sigma_e=t^2N_0\sum\limits_{j>0}\frac{|\langle\sigma|j\rangle|^2}{E-E_j}+\frac{|\langle\sigma|j^*\rangle|^2}{E+E_j}\\
\Sigma_h=t^2N_0\sum\limits_{j>0}\frac{|\langle\sigma|j^*\rangle|^2}{E-E_j}+\frac{|\langle\sigma|j\rangle|^2}{E+E_j}\\
\Sigma_A=-t^2N_0\sum\limits_{j>0}\frac{\langle\sigma|j\rangle\langle j|\sigma^*\rangle}{E-E_j}+\frac{\langle\sigma|j^*\rangle\langle j^*|\sigma^*\rangle}{E+E_j}\label{SigmaA}
\end{align}
Here $N_0=\mathrm{Im}\langle\theta|g^R_E|\theta\rangle $ is a normalization factor (for a point-contact $N_0=\pi\nu$). $\Sigma_A$ has a simple physical meaning: $2\Sigma_A$ is the amplitude of Andreev reflection (up to some phase factor) in the lowest order in $H_T$. Similarly, $\Sigma_{e,h}$ describe the corrections to electron-electron and hole-hole reflection due to the presence of $H_T$. Using Eq.\ \eqref{T1} we now construct the $\check{T}$-matrix:
\begin{align}
\check{T}=\check{G}_0\begin{pmatrix}|\theta\rangle &|\theta^*\rangle\end{pmatrix}\frac{N_0^{-1}\check{\Sigma}}{1-i\check{\Sigma}}\begin{pmatrix}\langle\theta|\\ \langle\theta^*|\end{pmatrix}
\label{Tsinglechannel}
\end{align}
where the $2\times2$ matrix in Eq.(\ref{Tsinglechannel}) reads as
\begin{multline}
\frac{\check{\Sigma}}{1-i\check{\Sigma}}=\frac{\check{\Sigma}+\check{1}(|\Sigma_A^2|-\Sigma_e\Sigma_h)}{(1-i\Sigma_e)(1-i\Sigma_h)+|\Sigma_A^2|}\label{SigmaGeometric}
\end{multline}
Finally we substitute $\check{G}=\check{G}_0+\check{T}\check{G}_0$ (and $\check{G}^A=\check{G}^{R\dagger}$) with $\check{T}$-matrix \eqref{Tsinglechannel} into the Kubo formula \eqref{Kubo}. Only off-diagonal (in Nambu space) terms of $\check{G}^{R,A}$ and consequently only off-diagonal terms of \eqref{SigmaGeometric} enter the trace, yielding
\begin{align}
G(eV)=\frac{2e^2}{h}\frac{4|\Sigma_A^2|}{\left(1-\Sigma_e\Sigma_h+|\Sigma_A^2|\right)^2+\left(\Sigma_e+\Sigma_h\right)^2}.\label{Gscml}
\end{align}
Voltage $eV$ enters the formula through $\check{\Sigma}$ where $E$ is set equal to $eV$. At resonance, $|eV-E_{j_0}|\ll W$, formula \eqref{Gscml} reproduces our single-level result \eqref{Gres} with $n_e=|\langle\sigma|j\rangle|^2t^2N_0/\pi$ and $n_h=|\langle\sigma|j^*\rangle|^2t^2N_0/\pi$. Away from any resonance, $\check{\Sigma}\ll1$, so that $G(eV)\simeq 2e^2/h\times4|\Sigma_A|^2$, which is the lowest order perturbative result in powers of $H_T$. $\Sigma_A$ is a sum of Andreev amplitudes over different levels $j$, and at some voltages several levels contribute to $\Sigma^A$ terms of comparable magnitudes, which may interfere. These interference effects are most drastic at lowest voltages, where interference pushes $G(0)$ to be equal either to $0$ or to $2e^2/h$, as we discuss below.

\subsection{Zero-bias conduction quantization}
The fact, that the zero-bias conductance of a single-channel NS-junction is quantized follows from general symmetry consideration \cite{Dahlhaus,PikulinNazarov}. The $S$-matrix describing an NS-junction has BdG-symmetry which relates $S^*_E$ and $S_{-E}$. In conjunction with unitarity $S^\dagger S=1$ this fixes $\det S_0=\pm1$, dividing possible $S$-matrices into two disconnected classes: trivial ($\det S_0=1$) and non-trivial ($\det S_0=-1$) .  The non-trivial case is represented by topological superconductor-metal junctions supporting a Majorana level on the junction \cite{OregOppen}, (this Majorana level is a well-defined quasistationary Andreev level if there is a barrier in the metal in the vicinity of the contact to superconductor, otherwise the Majorana level is leaked into the normal metal). In the single-channel case $S$ is a $2\times2$ matrix and $G(0)=(1-\det S_0)e^2/h$. Thus $G(0)$ is $0$ ($2e^2/h$) in the trivial (non-trivial) case. 

Let us now show that expression \eqref{Gscml} is indeed quantized at $E=0$. Suppose first there are no zero levels among $E_j$. Then, $\Sigma_A=0$ since terms with $|j\rangle $ and $|j^*\rangle$ in \eqref{SigmaA} cancel each other out at $E=0$, and consequently, $G(0)=0$. Suppose now that there is a single Majorana zero level $j_0$.
Then $|j_0\rangle=|j_0^*\rangle$ with $E_0=0$, leading to $\Sigma_e\simeq\Sigma_h\simeq|\Sigma_A|\sim 1/E$ and $\Sigma_e\Sigma_h-|\Sigma_A^2|=O(1)$ in the limit $E\to0$. Consequently, $G(0)=2e^2/h$ in this case.

The two above cases (no Majorana states, or a single one) describe all physical situations, 
as any higher number of zero modes would split. 
Let us show how it happens in somethat more details. One possible case corresponds to a non-Majorana zero level 
with $E_0=0$ and $|j_0\rangle\neq|j_0^*\rangle$, which can also be understood as a pair of uncoupled Majorana fermions $|j_0\rangle=e^{i\varphi}(|\gamma_1\rangle+i|\gamma_2\rangle)$. Depending on the choice of $\varphi$, the state $|j_0\rangle$ can be decomposed into different pairs of Majorana fermions $\gamma_1,\gamma_2$. If $|\langle\sigma| j_0\rangle|=|\langle\sigma| j_0^*\rangle|$ we choose $e^{2i\varphi}=\langle\sigma| j_0\rangle/\langle\sigma| j_0^*\rangle$ and find that the correspoding $\gamma_2$ is not coupled to the contact, $\langle\sigma|\gamma_2\rangle=0$. 
Thus, $\gamma_2$ does not enter the hamiltonian $H=H_T+H_M+H_S$  and we end up with the case of a single Majorana fermion $\gamma_1$. In the second case, $|\langle\sigma| j_0\rangle|\neq|\langle\sigma| j_0^*\rangle|$ both $\gamma_1$ and $\gamma_2$ enter $H_T$ for any choice of $\varphi$. The Majorana fermions are effectively coupled by the contact into a finite-energy fermionic level, which drives our NS-junction into the trivial regime. Indeed, if $|\langle\sigma| j_0\rangle|\neq|\langle\sigma| j_0^*\rangle|$ the $1/E^4$ singularities in the denominator of expression \eqref{Gscml} do not cancel out so that $G(0)=0$, as expected for a trivial junction. Finally, one may ask what happens if there are multiple zero levels $j_0,j_0'\dots$. This problem reduces to those considered above by changing the basis of the zero levels so that only one level couples to the contact, which is always possible for a single-channel contact. 

Note that exact quantization only happens in the single-channel case. For instance the exact cancellation of $|j\rangle $ and $|j^*\rangle$ Andreev amplitudes is only possible when the incoming electron and outgoing hole are in the same channel. This suggests that there may be some traces of the destructive interference for few-channel junctions, which become negligible as the number of channels becomes large. 

\subsection{Conductance at finite voltage  and interference}

The quantization discussed above only concerns $E=eV=0$ and cannot provide any information about finite energies. Our formula \eqref{Gscml} allows to find how narrow the quantization region is, i.e. at which voltages (or temperatures) it should be visible. The study of \eqref{Gscml} at finite $E$ also resolves an apparent paradox similar to that discussed in \cite{PikulinNazarov}: consider a finite superconducting system designed to host Majorana fermions. e.g. \cite{OregOppen,FuKane2008}. It is topologically trivial, because there is an even number of Majorana fermions across the system and there is finite energy splitting between them. So, $\det S_0=1$ and $G(0)=0$. Let us now extend the system to move one Majorana level to infinity. It is now uncoupled from the system and we are left with a topological junction which should show a zero-bias conductance peak. However, from continuity $S_0$ should remain in the trivial class and $G(0)$ should stay at zero. Below we show that both statements are true: there is a peak at low energies, 
and, at the same time, $G(0)=0$.

\begin{figure}
\includegraphics[width=1\columnwidth]{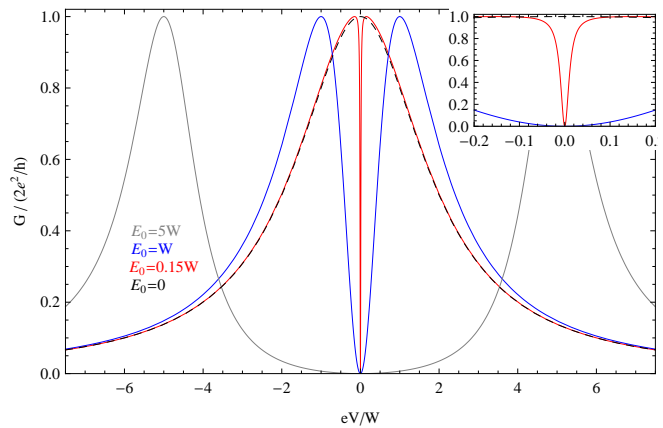}
\caption{(Color online) Single-channel conductance at $T=0$ for fixed $W$ and various values of $E_0$. At $E_0=5W$ (gray curve) there are two well-separated Lorentz peaks of width $W$ each. At $E_0=W$ (blue curve) the peaks are close to each other and significantly deformed. At $E_0=0.15W$ (red curve) the peaks merge into one Lorentz peak of width $2W$ with a narrow dip of the width $E_0^2/W$ in the middle. At $E_0=0$ (black dashed curve) there is no dip, and $G(eV)$ is an exact Lorentz peak of width $2W$. The inset shows the same picture at lowest voltages. $T_A^*$ is put to unity throughout our figures.}
\label{fig_fixedW}
\end{figure}

\begin{figure}
\includegraphics[width=1\columnwidth]{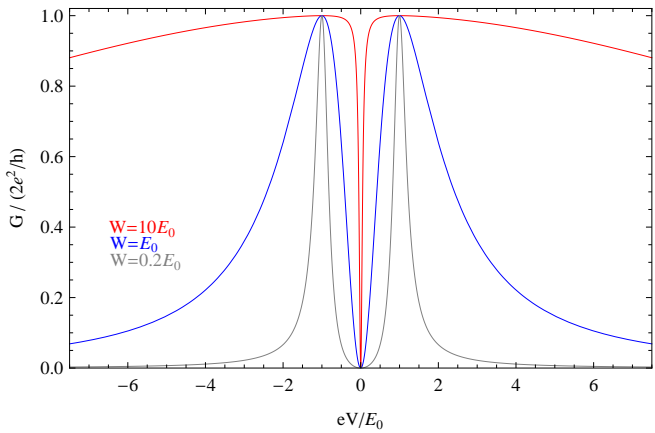}
\caption{(Color online) Single-channel conductance at $T=0$ for fixed $E_0$ and various values of $W$. At $W=0.2E_0$ (gray curve) there are two well-separated Lorentz peaks of width $W$ each. At $W=E_0$ (blue curve) the peaks are close to each other and significantly deformed. At $W=10E_0$ (red curve) the peaks merge into one Lorentz peak of width $2W$ with a narrow dip of the width $E_0^2/W$ in the middle.}
\label{fig_fixedE0}
\end{figure}

The low-voltage conductance is governed by the lowest energy level $|j\rangle  $ and $|j^*\rangle$ so we omit other levels in what follows -- they only lead to minor corrections in $G$. We find
\begin{align}
G(V)=\frac{2e^2}{h}\frac{1}{\left[\frac{e^2V^2-\tilde{E}_0^2}{2eVW}\right]^2+1}T_A^*\label{Gtwolevel}
\end{align}
with
\begin{align}
\tilde{E}_0^2=E_0^2+(|t_1|^2-|t_2|^2)^2\label{shift}\\
W=|t_1^2|+|t_2^2|\\
T_A^*=\frac{4|t_1t_2|^2}{\left(|t_1|^2+|t_2|^2\right)^2}\\
t_1=t\sqrt{N_0}\langle\sigma |j\rangle,\qquad\qquad t_2=t\sqrt{N_0}\langle\sigma |j^*\rangle
\end{align}
The shift in the resonance energy in \eqref{shift} is negligible, as it is smaller than $E_0$ by a factor of $W/\omega_0$ (except for the exotic case mentioned earlier when there are two uncoupled Majorana fermions in the superconducting system, which are then hybridized by the tunneling contact; in this case $\tilde{E}_0$ is the effective energy splitting and cannot be neglected). The two-level formula \eqref{Gtwolevel} agrees with what has been derived earlier in a different way for the special case when $H_T$ couples the metal tip to a single Majorana operator \cite{Flensberg}. If $E_0\gg W$ (gray curves on Fig.\ \ref{fig_fixedW},\ref{fig_fixedE0}), the function \eqref{Gtwolevel} shows two well-separated Lorentz peaks at $\pm E_0$ with a width $W$ each. If we decrease $E_0$, these two peaks move towards each other, but $G(0)$ stays zero. When $E_0$ becomes comparable to $W$ (blue curves on Fig.\ \ref{fig_fixedW},\ref{fig_fixedE0}) the shape does not look like a pair of Lorentz peak anymore. In the region $|eV|<E_0$ the conductance is suppressed due to destructive interference, while at $eV>|E_0|$ the conductance is enhanced: at $eV\gg E_0$ one gets $G(eV)=2e^2/h\cdot 4W^2/(eV)^2$ which is twice the sum of two Lorentz tails. Finally, at $E_0\ll W$ (red curves on Fig.\ \ref{fig_fixedW},\ref{fig_fixedE0}), we see one accurate Lorentz peak of double width $2W$ with a parametrically narrow dip in the region $E\lesssim E_0^2/W$ all the way to $0$. While quantization $G(0)=0$ is fulfilled, it only affects a very narrow region of energies. Note that throughout the change of $E_0$ the current $I(V\gg E_0,W)=\int_0^\infty G(V) dV$ (proportional to the area under the curves on Fig.\ \ref{fig_fixedW}) remains the same, and equals $2e/h\cdot\pi W$. 
Hence, at high temperatures $T\gg E_0,W$ the interference effect upon the conductance is  exponentially small.

\begin{figure}
\includegraphics[width=1\columnwidth]{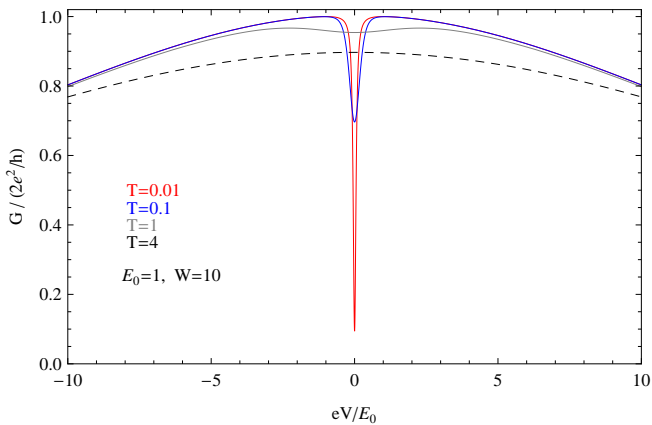}
\caption{(Color online) Single-channel conductance at $E_0=1$ and $W=10$ for different temperatures. At $T=0.01$ (red curve) there is a sharp dip with a finite but small $G(0)$. At $T=0.1$ (blue curve) which is of the order of the zero-temperature dip width $E_0^2/W$, the dip is significantly smeared out, while the wings of the Lorentz peak remain unchanged. At $T=1$ (gray curve) which is of the order of $E_0$ there is only a very shallow dip left. Finally, at $T=4$ (black dashed curve), which is comparable to $W=10$, there is no visible trace of the dip and the Lorentz peak starts to broaden.}
\label{fig_dipT}
\end{figure}

The above consideration solves the apparent paradox: the region of voltages and temperatures where the trivial character of the system is observable is very narrow, $eV,T\lesssim E_0$. Interestingly, increasing the coupling strength $W$ to the probe makes the observation of the conductance dip even more difficult -- at $W\gg E_0$ one needs to resolve energies of the order of $E_0^2/W$ instead of just $E_0$, as can be seen from Eq.\ \eqref{Gtwolevel} and Figs.\ \ref{fig_fixedW},\ref{fig_fixedE0},\ref{fig_dipT}. This can be understood in terms of the specific topological transition of the $S$-matrix described by Pikulin and Nazarov \cite{PikulinNazarov}. Our $S$-matrix has two poles (which are also poles of function \eqref{Gtwolevel}) at $E_\pm=\pm \sqrt{E_0^2-W^2}+iW$. For a weak contact (gray curve on Fig.\ \ref{fig_fixedE0}) they are connected by BdG-symmetry, $E_+=-E_-^*$ and represent a quasistationary Andreev level at $E_+$ and its image at $E_-$.
As $W$ is increased and the coupling gets stronger the poles move towards the imaginary axis. At $W=E_0$ (blue curve on Fig.\ \ref{fig_fixedE0}) they meet at the imaginary axis and the transition \cite{PikulinNazarov} happens. At higher $W$ one of the poles moves up the axis, the other down and each of them is now BdG-self-conjugate: $E_\pm=-E_\pm^*$. At very high $W$ (red curve on Fig.\ \ref{fig_fixedE0}) the lower pole is "buried"\cite{PikulinNazarov} -- it sits parametrically close to the real axis, $E_-=iE_0^2/2W$, and is very weakly coupled to the probe. This means, that the better the contact, the more our NS-junction looks topological from the electronic transport point of view. This is an important observation -- while the quantization of $G(0)$ happens for any $W$, a coupling stronger than the splitting $E_0$ diminishes  the temperature and voltage range in which the conductance dip (and hence the remote second Majorana fermion) is observable, as shown in Fig.\ \ref{fig_dipT}.

\section{Discussion}
\label{sec:discussion}

We now  discuss real systems where the  effects studied can be observed. A natural object to look at is 
a core of an Abrikosov vortex in a quasi-2D  superconductor. 
The  above results are applicable if the Caroli-De Gennes-Matricon (CdGM) levels~\cite{CdGM,deGennes} accessible by the tunnel contact on the top plane of the superconductor \textbf{are not} spreaded into the continuous band dispersing 
along the vortex core.
The CdGM spectrum of a clean vortex core is described by
\begin{align}
E_{\mu,p_z}=\mu\omega_0  + \mu\frac{p_z^2}{2m^*}
\label{disc1}
\end{align}
with
$
\omega_0\sim\frac{\Delta}{p_F\xi}$ and
 $ m^*= {m_z}(p_{F\perp}\xi)^2 $,
where $\mu$ is the angular momentum, $\xi$ is the coherence length and $m_z$ is the effective 
mass along the $z$-direction assumed to be larger than $(x,y)$-plane  effective mass $m$ in an anisotropic superconductor. In the presence of even a low concentration of impurities, CdGM states in a neighboring atomic planes
of layered material will be displaced by the amount of energy $\sim \omega_0$, whereas  bandwidth of each level
is about $\mu \omega_0 m/m_z$.  Therefore, in strongly anisotropic materials with $m_z \gg m$  individual low-lying CdGM levels are well localized ~\cite{Skvor1,Skvor2}  and tunnelling from the STM tip into the surface of such a material may show the features discussed above in this paper. However, in order to discern these features, one needs to work at very low
temperatures  $ T \leq \omega_0$.  Currently, it should be possible to reach this condition
 while dealing with a material like  NbSe$_2$, whereas older data~\cite{NbSe2,NbSe2prime} were taken at too high temperatures.

Another possibility is to use genuine two-dimensional electron systems like graphene or a surface of topolgical
insulator. In these materials there are no transverse quantum numbers and the Andreev spectrum is discrete with level spacing $\omega_0$ (in graphene the spectrum is fourfold degenerate due to valley and spin degrees of freedom).
An additional advantage of these two-dimensional systems is a controllable $E_f$. The charge neutrality point in graphene coincides with the $K$-point of its Dirac spectrum, which allows to have $\omega_0$ comparable to $E_f$.
 An important issue is the implementation of superconductivity in these two-dimensional systems. A possible solution is to cover them with a conventional superconductor to induce superconductivity by proximity effect \cite{ProxGraphene1,ProxGraphene2,FuKane2008,Sacepe}. However, one needs to make sure that there are no parasite Andreev states in the covering superconductor. This can be done by making a hole in the superconductor of a radius of the order of the coherence length \cite{Nori,IOF,Melnikov} physically removing the normal vortex core with all its low-lying Andreev states from the superconductor. In addition, such a hole should facilitate the application of a tunneling microscope to the graphene or topological insulator surface.

\section{Conclusion}
In conclusion, we have studied resonant Andreev reflection in an arbitrary tunneling N-S contact between a metal and a superconductor hosting a discrete Andreev spectrum $E_j$. If the level spacing is larger than the inverse dwell time $W$, the subgap conductance $G(V)$ consists of Lorentz peaks centered at $eV=E_j$. We have derived formulae describing these peaks in the general case. The heights of the peaks do not scale with the coupling strength and are generally very close to $2e^2/h$, while widths scale as normal conductance of the contact. We also established an exact formula describing the Andreev conductance for a generic single-channel contact that takes into account all Andreev levels at once. This formula reveals the role of interference between Andreev reflection processes involving different levels $E_j$. At $eV=0$ interference between the level $E_j$ and its particle-hole image $-E_j$ is destructive, so that $G(0)$ is exactly $2e^2$ or $0$ depending on whether there is a Majorana fermion in the spectrum, in agreement with $S$-matrix theory of topological NS-junctions. We have also studied $G(V)$ for finite energies for different $W$ and lowest energy level $E_0$ and explained what happens when the system is gradually driven from the trivial to the topological phase by removing a Majorana fermion from the system and thus driving $E_0$ to zero.  We have found that at $E_0\ll W$ the conductance shows a zero-bias Lorentz peak of the usual width $2W$, with a very narrow dip on top of it, which secures $G(0)=0$. The parametrically small width $E_0^2/W$ of the dip can be understood as the "burying" of the remote Majorana fermion: the strong coupling of the tunneling tip to one of the Majorana states effectively decreases the coupling of the system to the other Majorana state. To observe the dip (i.e. the "buried Majorana fermion") experimentally, temperature and energy resolution has to be smaller than $E_0^2/W$. The effects we discussed should be observable in two-dimensional electronic systems including graphene or the surface of three-dimensional topological insulators with proximity-induced superconductivity. Another possibility is
to study the level structure of  Abrikosov vortices in a layered strongly anisotropic superconductor like NbSe$_2$ at ultra-low temperatures.

We thank  Alexey Ioselevich, Pavel Ostrovsky and Dmitry Pikulin for very helpful discussions.
This work was supported by the  RFBR grant \# 10-02-00554-a and by the RAS program "Mesoscopic and disordered systems".

\end{document}